\documentclass[runningheads]{llncs}
\usepackage[T1]{fontenc}
\usepackage{graphicx}
\usepackage{amsmath}
\usepackage{amssymb}
\usepackage{amsfonts}
\usepackage{mathtools}

\usepackage{booktabs}       
\usepackage{multirow}       
\usepackage{array}          
\usepackage{graphicx}       
\usepackage{adjustbox}      

\usepackage{tikz}           
\usepackage{float}          
\usepackage{subfigure}      
\usepackage{caption}

\usepackage{algorithm}
\usepackage{algorithmic}

\usepackage{xcolor}         
\usepackage{url}            
\usepackage{enumitem}       
\usepackage{bbm}            
\usepackage{bm}             
\usepackage{dsfont}         
\usepackage{hyperref}       

\usepackage{cleveref}       
\usepackage{amssymb}
\usepackage[utf8]{inputenc}
\usepackage{booktabs}
\usepackage{multirow}
\usepackage{xcolor}
\usepackage{colortbl}
\usepackage{makecell} 
\usepackage{tabularx}
\usepackage{array}
\usepackage{orcidlink}

\usepackage{marvosym}

\begin{document}
\title{Prompt-Aware Adaptive Elastic Weight Consolidation for Continual Learning in Medical Vision-Language Models}
\titlerunning{PA-EWC for Continual Learning in Medical Vision-Language Models}
%

\author{
  Ziyuan Gao\textsuperscript{(\Letter)}\orcidlink{0009-0005-8092-5576} \and
  Philippe Morel\orcidlink{0000-0002-9847-3409}
}

\authorrunning{Z. Gao and P. Morel}

\institute{
  University College London\\
  \email{clairegao0930@gmail.com}
}

\maketitle
\begin{abstract}
Medical AI systems face catastrophic forgetting when deployed in clinical settings, where models must learn new imaging protocols while retaining prior diagnostic capabilities. This challenge is particularly acute for medical vision-language models that must preserve complex cross-modal alignments between medical images and clinical terminology across diverse imaging modalities. We introduce Prompt-Aware Adaptive Elastic Weight Consolidation (PA-EWC), a novel continual learning approach that addresses catastrophic forgetting through prompt-guided parameter specialization. Our method systematically categorizes model parameters based on their functional roles in processing visual-descriptive, spatial-guided, and medical-semantic information, enabling targeted protection of critical knowledge while allowing adaptation to new clinical requirements. PA-EWC incorporates adaptive Fisher Information computation with gradient stability analysis and develops weighted complexity metrics based on medical terminology density. We evaluate our approach across five medical imaging datasets (Kvasir-SEG, ISIC 2018, CheXlocalize, BUSI, CAMUS) representing diverse modalities including endoscopy, dermoscopy, radiography, and ultrasound. 
Experimental results demonstrate that PA-EWC reduces catastrophic forgetting by up to 17.58\% compared to baseline methods, with performance improvements of 4.30\% on chest X-ray pathology localization and 6.06\% on polyp segmentation.


\end{abstract}

\section{Introduction}

In medical artificial intelligence (AI) systems, a significant challenge arises when these systems are deployed in clinical settings: \textbf{catastrophic forgetting}. As medical institutions acquire new imaging protocols and encounter novel pathological conditions, AI models must learn these new tasks while retaining their prior diagnostic proficiencies. However, when neural networks learn new medical tasks sequentially, they often experience a drastic decline in performance on previously learned tasks~\cite{kirkpatrick2017overcoming}. This issue is particularly pronounced in medical vision-language models, which are integral to modern diagnostic workflows and must preserve complex cross-modal alignments between medical images and clinical terminology. Unlike models trained on natural image domains, medical imaging presents distinct challenges, including diverse imaging modalities (e.g., endoscopy, dermoscopy, radiography) and strict safety requirements where the loss of diagnostic capabilities could compromise patient care.

Medical AI systems deployed in clinical practice face significant limitations when adapting to new requirements. Production systems for diabetic retinopathy screening~\cite{gulshan2016development} and radiology workflows~\cite{mckinney2020international} typically require complete retraining when integrating new imaging protocols. This leads to substantial downtime and computational costs that disrupt patient care workflows. While continual learning approaches like Elastic Weight Consolidation (EWC)~\cite{kirkpatrick2017overcoming} have shown promise in research settings, they apply uniform parameter protection across all model components. This proves inadequate for medical domains, which require specialized parameter handling for diverse clinical tasks. The core challenge remains: how can medical vision-language models continuously adapt to evolving clinical needs without losing diagnostic capabilities?

We introduce \textbf{Prompt-Aware Adaptive Elastic Weight Consolidation (PA-EWC)}, a novel approach that addresses catastrophic forgetting in medical vision-language models through prompt-guided parameter specialization. By selectively protecting parameters based on their functional importance to specific medical tasks, PA-EWC preserves critical knowledge while allowing adaptation to new clinical requirements. Our method leverages medical language complexity analysis and gradient-based parameter classification to enable targeted protection of functionally specialized components.

Our contributions are fourfold: (1) \textbf{Prompt-Guided Parameter Classification} that systematically categorizes model parameters based on their functional specialization in processing different types of medical language; (2) \textbf{Adaptive Fisher Information with Gradient Stability} that incorporates gradient stability analysis and task similarity measures for dynamic importance estimation; (3) \textbf{Parameter-Aligned Complexity Assessment} that develops weighted complexity metrics based on medical terminology density; and (4) \textbf{Dynamic Multi-Group Protection} that selectively protects three distinct parameter groups based on task-specific linguistic patterns. 
Experiments across five medical imaging datasets (Kvasir-SEG~\cite{jha2020kvasir} for polyp segmentation, ISIC 2018~\cite{codella2019skin} for skin lesions, CheXlocalize~\cite{selvan2020chexlocalize} for chest X-rays, BUSI~\cite{al2020dataset} for breast ultrasound, CAMUS~\cite{leclerc2019deep} for cardiac ultrasound) show our PA-EWC method reduces catastrophic forgetting by up to 17.58\%. Performance improves by 4.30\% on challenging chest X-ray pathology localization and by 6.06\% on polyp segmentation compared to naive sequential learning.

\section{Related Work}

\subsection{Continual Learning in Medical Vision-Language Models}
Continual learning has emerged as a critical challenge in medical AI systems, where models must adapt to new clinical protocols while preserving existing diagnostic capabilities. Traditional approaches to catastrophic forgetting, such as Elastic Weight Consolidation (EWC)\cite{kirkpatrick2017overcoming}, Progressive Neural Networks\cite{rusu2016progressive}, and PackNet~\cite{mallya2018packnet}, have shown promise in general computer vision tasks but face unique challenges in medical domains due to the heterogeneous nature of medical imaging modalities. Recent advances in prompt-based continual learning, including Learning to Prompt (L2P)\cite{wang2022learning}, Visual Prompt Tuning (VPT)\cite{jia2022visual}, DualPrompt~\cite{wang2022dualprompt}, and CODA-Prompt~\cite{smith2023coda}, have introduced more sophisticated approaches but primarily focus on natural image domains and fail to address the specialized requirements of medical multimodal understanding. The medical domain presents unique challenges including diverse imaging modalities (endoscopy, dermoscopy, radiography, ultrasound) with distinct visual characteristics and specialized vocabularies~\cite{esteva2017dermatologist}, clinical deployment requirements that demand high reliability~\cite{rajpurkar2017chexnet}, and the need for precise alignment between clinical terminology and anatomical structures~\cite{gulshan2016development,mckinney2020international}. 
Recent work in medical continual learning has begun to address these domain-specific challenges through medical knowledge-aware regularization, but still applies uniform protection across model components without considering functional specialization in multimodal processing.

\subsection{Adaptive Parameter Management in Continual Learning}

The effectiveness of continual learning methods critically depends on accurately identifying which parameters are important for previous tasks and should be protected from modification. Several approaches have attempted to improve parameter importance estimation, including Synaptic Intelligence~\cite{zenke2017continual} which tracks parameter importance through online estimation, Memory Aware Synapses~\cite{aljundi2018memory} which accumulates importance weights across tasks, and Gradient Episodic Memory~\cite{lopez2017gradient} which stores representative examples from previous tasks. Adaptive methods like Self-Adaptive EWC have introduced dynamic importance weighting based on task similarity measures, while functional regularization approaches~\cite{titsias2019functional} focus on preserving input-output mappings rather than specific parameter values, and meta-learning frameworks~\cite{finn2017model} attempt to find initialization points that facilitate rapid adaptation. 
However, these approaches have not been systematically applied to medical vision-language models, which require specialized parameter protection strategies based on task-specific linguistic patterns.

\section{Methodology}

\subsection{Problem Formulation}

In continual learning for medical multimodal vision-language models, we address the catastrophic forgetting problem when sequentially learning tasks $\mathcal{T}_1, \mathcal{T}_2, \ldots, \mathcal{T}_N$. Each task $\mathcal{T}_i$ consists of medical images $\mathcal{X}_i$, corresponding segmentation masks $\mathcal{Y}_i$, and textual prompts $\mathcal{P}_i$ with varying complexity levels. The model parameters $\theta$ must be updated to optimize performance on the current task while preserving knowledge from previous tasks.



\begin{figure}[H]
\centering
\includegraphics[width=\textwidth]{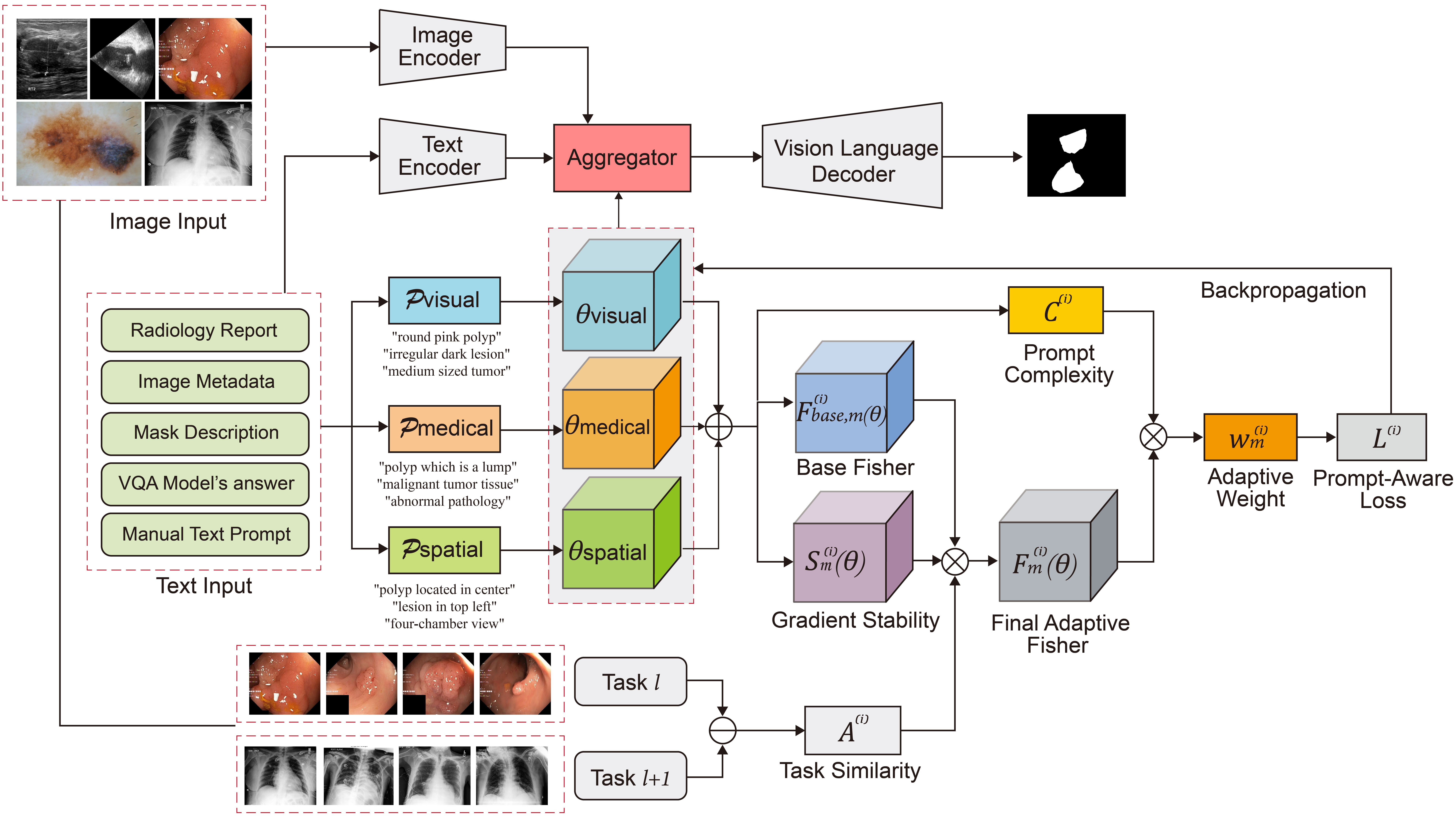}
\caption{Overall PA-EWC pipeline}
\label{fig:organ}
\end{figure}

For a model $f_\theta$ with parameters $\theta$, we aim to minimize performance degradation on previous tasks while maximizing adaptation to new tasks:
\begin{equation}
\min_\theta \sum_{i=1}^{N} \mathcal{L}_i(f_\theta(\mathcal{X}_i, \mathcal{P}_i), \mathcal{Y}_i) \text{ subject to } \forall j < i: \mathcal{L}_j(f_\theta) \leq \mathcal{L}_j(f_{\theta_j^*})
\end{equation}
where $\theta_j^*$ represents the optimal parameters after training on task $j$.

\subsection{Prompt-Guided Parameter Classification}

 We propose a prompt-aware parameter classification strategy that selectively protects parameters based on their functional specialization. Our approach consists of two components: (1) a hierarchical prompt taxonomy that categorizes medical language complexity, and (2) an empirical parameter classification method that maps model components to prompt response patterns. This enables task-specific protection while allowing non-essential parameters to adapt to new tasks.

To systematically analyze parameter-prompt relationships, we establish a five-tier prompt hierarchy (Table \ref{tab:prompt_hierarchy}) with three core functional categories: Visual-Descriptive prompts ($\mathcal{P}_{\text{visual}}$) emphasizing visual attributes, Spatial-Guided prompts ($\mathcal{P}_{\text{spatial}}$) incorporating anatomical positioning, and Medical-Semantic prompts ($\mathcal{P}_{\text{medical}}$) providing clinical definitions. To validate the effectiveness of our approach, we design two additional categories: basic and comprehensive prompts to serve as baselines, as demonstrated in the experimental sections.

\begin{center}
\fontsize{8}{10}\selectfont
\captionof{table}{Hierarchical Prompt Strategy}
\label{tab:prompt_hierarchy}
\begin{tabularx}{\textwidth}{>{\raggedright\arraybackslash}p{0.16\textwidth}>{\raggedright\arraybackslash}p{0.56\textwidth}>{\raggedright\arraybackslash}p{0.26\textwidth}}
\toprule
\textbf{Type} & \textbf{Prompt Examples} & \textbf{Key Features} \\
\midrule
\cellcolor{yellow!30}Basic & 
\cellcolor{yellow!15}"polyp", "skin lesion", "foot ulcer", "myocardium" & 
Simple medical terminology \\
\midrule
\cellcolor{cyan!30}Visual-Descriptive \newline ($\mathcal{P}_{\text{visual}}$) & 
\cellcolor{cyan!15}"round polyp", "pink round polyp", "medium pink round polyp", "irregular skin melanoma", "one small irregular foot ulcer" & 
Visual attributes, shape descriptions, size modifiers \\
\midrule
\cellcolor{lime!30}Spatial-Guided \newline ($\mathcal{P}_{\text{spatial}}$) & 
\cellcolor{lime!15}"polyp located in center", "skin melanoma located in top left", "foot ulcer located in bottom right", "myocardium located in left ventricle" & 
Location guidance, spatial references, anatomical positioning \\
\midrule
\cellcolor{orange!30}Medical-Semantic \newline ($\mathcal{P}_{\text{medical}}$) & 
\cellcolor{orange!15}"polyp which is a small lump in colon", "skin lesion which is an abnormal tissue growth", "melanoma which is a malignant tumor" & 
Medical definitions, clinical descriptions, pathological context \\
\midrule
\cellcolor{pink!30}Comprehensive & 
\cellcolor{pink!15}"medium pink round polyp which is a small lump in colon located in top left", "irregular skin melanoma which is a malignant tumor located in center" & 
Complete integration of all information types, maximum complexity \\
\bottomrule
\end{tabularx}
\end{center}

Using these prompt types, we empirically determine parameter-prompt relationships through controlled gradient analysis. For each core prompt category, we measure parameter gradient response magnitude and assign parameters to the group yielding maximum response:
\begin{equation}
\text{class}(\theta_k) = \arg\max_{t} \left|\nabla_{\theta_k} \mathcal{L}(f_{\theta}(x, \mathcal{P}_t), y)\right|_2
\end{equation}

where $R_k^{(t)}$ represents the gradient response magnitude of parameter $\theta_k$ to prompt type $t$, $\mathcal{L}$ is the loss function, and $\mathcal{P}_t$ denotes the prompt template for task type $t$.
This analysis reveals three parameter groups (Table \ref{tab:parameter_classification}): Visual-Descriptive Features ($\theta_{\text{visual}}$) mapping to vision processing layers, Spatial-Guided Attention ($\theta_{\text{spatial}}$) corresponding to cross-attention modules, and Medical-Semantic Understanding ($\theta_{\text{medical}}$) associated with text components.

This classification enables task-adaptive parameter protection: when a new task exhibits specific linguistic patterns (spatial, visual, or medical), we selectively strengthen protection for the corresponding parameter groups. This balances knowledge retention and adaptation flexibility in continual learning.

\begin{center}
\captionof{table}{Parameter Classification Based on Prompt Response Patterns}
\label{tab:parameter_classification}
{\fontsize{8}{10}\selectfont
\begin{tabularx}{\textwidth}{>{\raggedright\arraybackslash}p{0.2\textwidth}>{\raggedright\arraybackslash}p{0.2\textwidth}>{\raggedright\arraybackslash}p{0.35\textwidth}>{\raggedright\arraybackslash}p{0.24\textwidth}}
\toprule
\textbf{Parameter Group} & \textbf{Primary Response to} & \textbf{Key Vocabulary} & \textbf{Model Components} \\
\midrule
\cellcolor{cyan!20}Visual-Descriptive \newline Features \newline $(\theta_{\text{visual}})$ & 
Visual attribute descriptions \newline ($\mathcal{P}_{\text{visual}}$) & 
\texttt{round, irregular, pink, medium, small, large, shape, color, texture} & 
Vision processing layers, feature extraction modules \\
\midrule
\cellcolor{lime!20}Spatial-Guided \newline Attention \newline $(\theta_{\text{spatial}})$ & 
Spatial guidance prompts \newline ($\mathcal{P}_{\text{spatial}}$) & 
\texttt{located, center, left, right, top, bottom, four-chamber, two-chamber} & 
Cross-attention layers, conditional processing modules \\
\midrule
\cellcolor{orange!20}Medical-Semantic \newline Understanding \newline $(\theta_{\text{medical}})$ & 
Medical terminology and semantics \newline ($\mathcal{P}_{\text{medical}}$) & 
\texttt{polyp, lesion, tumor, lump, pathology, malignant, benign, tissue} & 
Text processing layers, medical vocabulary modules \\
\bottomrule
\end{tabularx}
}
\end{center}

\subsection{Parameter-Aligned Prompt Complexity Assessment}

To quantify prompt complexity and guide adaptive parameter protection, we develop a weighted complexity metric that leverages our established parameter groups ($\theta_{\text{visual}}$, $\theta_{\text{spatial}}$, $\theta_{\text{medical}}$). Rather than using traditional word-count or syntactic measures, our approach assigns differential weights to vocabulary types based on their functional specialization in medical multimodal understanding.

The complexity calculation uses the vocabulary categories from Table \ref{tab:parameter_classification}:
\begin{equation}
C^{(i)} = \frac{1}{|\mathcal{P}^{(i)}|} \sum_{p \in \mathcal{P}^{(i)}} \left(|W_p| + \alpha_{\text{visual}}|V_p| + \alpha_{\text{spatial}}|S_p| + \alpha_{\text{medical}}|M_p|\right)
\end{equation}
where $|W_p|$, $|V_p|$, $|S_p|$, and $|M_p|$ represent base words, visual descriptors, spatial terms, and medical terms in prompt $p$, respectively. The weighting factors $\alpha_{\text{visual}} = 2.0$, $\alpha_{\text{spatial}} = 2.5$, and $\alpha_{\text{medical}} = 3.0$ emphasize domain-specific terminology. Prompts with higher concentrations of specialized medical vocabulary thus receive proportionally stronger parameter protection.

\subsection{Adaptive Fisher Information Matrix with Gradient Stability}

Traditional Fisher Information matrices in EWC assume uniform parameter importance across all learning phases, which fails to capture the dynamic nature of parameter significance in continual learning scenarios. We enhance the Fisher computation by incorporating gradient stability analysis and task similarity measures to provide more accurate importance estimation for each parameter group.

For each parameter group $m \in \{\text{visual}, \text{spatial}, \text{medical}\}$, we compute adaptive Fisher Information that incorporates gradient stability and task similarity. The base Fisher computation follows the standard formulation:
\begin{equation}
F_{\text{base},m}^{(i)}(\theta) = \mathbb{E}_{x \sim \mathcal{D}_i} \left[ \left(\nabla_{\theta_m} \log p(y|x; \theta_m)\right)^2 \right]
\end{equation}

We augment this base computation with two adaptive factors. The gradient stability factor $S_m^{(i)}(\theta) = \sigma(-\text{Var}(\nabla_{\theta_m}))$ measures parameter consistency across training samples, where higher variance indicates unstable gradients that should receive reduced protection. The task similarity factor captures activation pattern changes between consecutive tasks:
\begin{equation}
A^{(i)} = \frac{1}{L} \sum_{l=1}^{L} \frac{1}{1 + |\mu_l^{(i-1)} - \mu_l^{(i)}| + |\sigma_l^{(i-1)} - \sigma_l^{(i)}|}
\end{equation}
where $\mu_l$ and $\sigma_l$ represent the mean and standard deviation of activations in layer $l$. This similarity measure approaches its maximum value for highly similar tasks and approaches 1.0 for dissimilar tasks, allowing stronger protection when tasks share common activation patterns.

The final adaptive Fisher matrix combines these factors through weighted multiplication:
\begin{equation}
F_m^{(i)}(\theta) = F_{\text{base},m}^{(i)}(\theta) \cdot S_m^{(i)}(\theta) \cdot A^{(i)}
\end{equation}
This formulation ensures that parameters with stable gradients and high task similarity receive maximum protection, while unstable or task-specific parameters are allowed greater adaptation flexibility.

\subsection{Adaptive Weight Computation}
We dynamically adjust protection weights for each parameter group based on prompt complexity and task similarity. For each parameter group $m \in \{\text{visual},$ $\text{spatial}, \text{medical}\}$, the adaptive weight is computed as:
\begin{equation}
w_m^{(i)} = F_m^{(i)}(\theta) \times \left(1 + \frac{C^{(i)}}{C_{\text{max}}}\right)
\end{equation}
where $F_m^{(i)}(\theta)$ is the adaptive Fisher Information matrix from Equation (6), $C^{(i)}$ is the prompt complexity for task $i$, and $C_{\text{max}}$ is the maximum observed complexity across all tasks.

This formulation increases protection weights when prompts are more complex, while maintaining the gradient stability and task similarity properties established in the Fisher Information computation.

\subsection{Prompt-Aware Loss Function}

The proposed framework employs a composite loss function that balances current task performance with knowledge retention from previous tasks. The total loss function integrates segmentation objectives with adaptive continual learning constraints:
\begin{equation}
\mathcal{L}_{\text{total}}^{(i)} = w_{\text{seg}} \mathcal{L}_{\text{seg}}^{(i)} + w_{\text{dice}} \mathcal{L}_{\text{dice}}^{(i)} + w_{\text{ewc}} \sum_{j=1}^{i-1} \mathcal{L}_{\text{EWC}}^{(j)}
\end{equation}
The enhanced EWC component accumulates regularization constraints from all previous tasks, weighted by parameter-specific importance across visual, spatial, and medical parameter groups:
\begin{equation}
\mathcal{L}_{\text{EWC}}^{(j)} = \sum_{m} w_m^{(j)} \sum_{\theta \in \theta_m} F_m^{(j)}(\theta) \cdot (\theta - \theta_m^{(j)*})^2
\end{equation}
where $m$ denotes the parameter group (visual, spatial, or medical), $w_m^{(j)}$ represents the adaptive weights from Equation (8), $F_m^{(j)}(\theta)$ denotes the Fisher Information importance for parameter $\theta$ in group $m$ at task $j$, and $\theta_m^{(j)*}$ are the optimal parameters for group $m$ after completing task $j$.

The Dice loss component addresses class imbalance inherent in medical segmentation tasks through soft probability weighting:
\begin{equation}
\mathcal{L}_{\text{dice}}^{(i)} = 1 - \frac{1}{B} \sum_{b=1}^{B} \frac{2 \cdot \sum_{h,w} p_{b,h,w}^{(fg)} \cdot y_{b,h,w}}{\sum_{h,w} p_{b,h,w}^{(fg)} + \sum_{h,w} y_{b,h,w} + \epsilon}
\end{equation}
where $p^{(fg)} = \text{softmax}(\text{logits})[:, 1]$ represents the foreground probability predictions. The loss weighting scheme employs $w_{\text{seg}} = 1.0$, $w_{\text{dice}} = 0.3$, and $w_{\text{ewc}} = 10.0$ to prioritize knowledge retention while maintaining task-specific performance, with base Fisher Information importance scaled to 1000.

\section{Experiment}

\subsection{Datasets and Evaluation Metric}


We conduct comprehensive experiments across five major medical imaging datasets to evaluate continual learning performance in diverse clinical scenarios. Our evaluation encompasses \textbf{Kvasir-SEG} \cite{jha2020kvasir} for polyp segmentation (800/100/100 train/val/test), \textbf{ISIC 2018} \cite{codella2019skin} for skin lesion segmentation (810/90/379), \textbf{CheXlocalize} \cite{selvan2020chexlocalize} for chest X-ray pathology localization (1279/446/452), \textbf{BUSI} \cite{al2020dataset} for breast ultrasound tumor segmentation (624/78/78), and \textbf{CAMUS} \cite{leclerc2019deep} for cardiac ultrasound chamber segmentation (4800/600/600).
These datasets represent distinct imaging modalities (endoscopy, dermoscopy, radiography, ultrasound), anatomical regions (gastrointestinal, dermatological, pulmonary, mammographic, cardiac), and segmentation complexities, providing a comprehensive evaluation framework for medical vision-language continual learning.

We assess continual learning performance using two primary metrics: 

(1) \textbf{Dice Coefficient} measuring segmentation accuracy across all tasks, computed as 
\begin{equation}
\text{Dice}_i^{final} = \frac{2|P_i \cap G_i|}{|P_i| + |G_i|}
\end{equation}
with $P_i$ and $G_i$ being the predicted and ground truth segmentation masks for task $i$, and $T$ is the total number of tasks; 

(2) \textbf{Forgetting Rate} \cite{kirkpatrick2017overcoming} quantifying knowledge retention, calculated as 
\begin{equation}
\text{FR} = \sum_{i=1}^{T-1} (\text{Dice}_i^{peak} - \text{Dice}_i^{final})
\end{equation}
where $\text{Dice}_i^{peak}$ is the peak performance on task $i$ and $\text{Dice}_i^{final}$ is the final performance after learning all subsequent tasks. Lower forgetting rates indicate superior catastrophic forgetting mitigation. 
Additionally, we report training efficiency metrics including computational time and parameter overhead to demonstrate practical deployment feasibility.

\subsection{Implementation Details}

 We employ CLIPSeg \cite{luddecke2022image} as the base vision-language segmentation model, featuring a ViT-B/16 vision encoder and transformer-based text encoder with cross-modal attention mechanisms for processing 512$\times$512 resolution images with text prompts. All experiments utilize AdamW optimizer with learning rate 1e-4, weight decay 1e-2, and cosine annealing scheduler, where sequential training employs 50 epochs per task with early stopping based on validation Dice coefficient. Training is conducted with batch size 16 on 2×A100 GPUs. Results are averaged over three runs with random seed 43.

\subsection{Comparison with State-of-the-Art Methods}


To demonstrate the effectiveness of our approach, we compare against recent state-of-the-art continual learning methods adapted for medical vision-language tasks. Table~\ref{tab:sota_comparison} presents comprehensive comparisons across classic continual learning, parameter-efficient fine-tuning, and vision-language model specific methods.
PA-EWC achieves state-of-the-art performance, outperforming the strongest baseline (ZSCL) by 2.32\% in Dice coefficient and reducing catastrophic forgetting by 2.45\%, while maintaining superior computational efficiency with 8.7-hour training time compared to ZSCL's 9.1 hours. The results demonstrate consistent improvements across all evaluation metrics, with our method achieving 75.34\% average Dice score and only 18.42\% forgetting rate. With 155M parameters, PA-EWC delivers the best performance across all metrics while maintaining balanced training efficiency, establishing its practical viability.

\begin{table}[htbp]
\fontsize{8}{9}\selectfont
\centering
\caption{Comparison with state-of-the-art continual learning methods on medical vision-language segmentation. Training time measured on 2×A100 GPUs. Best results in \textbf{bold}, second best \underline{underlined}.}
\label{tab:sota_comparison}
\begin{tabular*}{\textwidth}{@{\extracolsep{\fill}}lcccc@{}}
\toprule
\textbf{Method} & \textbf{\makecell{Average \\ Dice (\%)}} & \textbf{\makecell{Average \\ Forgetting (\%)}} & \textbf{\makecell{Training \\ Time (h)}} & \textbf{\makecell{Params \\ (M)}} \\
\midrule
\multicolumn{5}{l}{\textit{Basic Baselines}} \\
CLIPSeg Individual & 78.5$\pm$0.6 & 0.0$\pm$0.0 & 12.4 & 150.0 \\
CLIPSeg Sequential & 62.3$\pm$1.9 & 35.7$\pm$2.3 & 8.1 & 150.0 \\
\midrule
\multicolumn{5}{l}{\textit{Classic Continual Learning Methods}} \\
EWC~\cite{kirkpatrick2017overcoming} (2017) & 65.48$\pm$1.7 & 28.34$\pm$2.1 & 8.5 & 150.0 \\
LwF~\cite{li2017learning} (2017) & 67.2$\pm$1.5 & 26.8$\pm$1.9 & 8.9 & 150.0 \\
PackNet~\cite{mallya2018packnet} (2018) & 68.92$\pm$1.4 & 24.71$\pm$1.8 & 9.2 & 150.0 \\
Progressive NN~\cite{rusu2016progressive} (2016) & 71.23$\pm$1.2 & 22.15$\pm$1.5 & 15.3 & 375.0 \\
DER++~\cite{buzzega2020dark} (2020) & 71.58$\pm$1.0 & 22.43$\pm$1.3 & 9.8 & 155.0 \\
\midrule
\multicolumn{5}{l}{\textit{Parameter-Efficient Fine-tuning Methods}} \\
CLIPSeg + LoRA~\cite{hu2021lora} (2021) & 69.5$\pm$1.2 & 24.3$\pm$1.6 & \textbf{6.8} & \underline{152.5} \\
CLIPSeg + CLIP-Adapter~\cite{gao2021clip} (2021) & 70.1$\pm$1.3 & 23.7$\pm$1.4 & \underline{6.9} & \textbf{151.8} \\
\midrule

ZSCL~\cite{zheng2023preventing} (2023) & \underline{73.02$\pm$0.9} & \underline{20.87$\pm$1.2} & 9.1 & 150.0 \\
MoE-Adapters~\cite{yu2024boosting} (2024) & 72.84$\pm$1.1 & 21.34$\pm$1.4 & 9.4 & 165.3 \\
\midrule
PA-EWC (Ours) & \textbf{75.34$\pm$0.8} & \textbf{18.42$\pm$1.0} & 8.7 & 155.0 \\
\bottomrule
\end{tabular*}
\end{table}

\subsection{Forgetting Rate Analysis}

To further assess catastrophic forgetting behavior, we analyze forgetting rates across different task orders using five distinct sequential learning scenarios. Table~\ref{tab:forgetting_comparison} compares forgetting rates between sequential baseline and our PA-EWC method across all five medical imaging datasets.
Our PA-EWC method demonstrates superior forgetting mitigation compared to sequential training in most scenarios. The approach achieves significant improvements in challenging cases, reducing forgetting from 47.56\% to 31.50\% for CAMUS and from 49.32\% to 31.74\% for CheX in the ISIC→CAMUS→Kvasir→CheX→BUSI sequence. The color-coded visualization shows task learning order, with earlier tasks generally experiencing higher forgetting rates, confirming the expected catastrophic forgetting pattern in continual learning scenarios. PA-EWC consistently outperforms sequential training across different task orderings, demonstrating robust and reliable performance regardless of learning sequence variations.

\definecolor{depth1}{RGB}{180,100,100}    
\definecolor{depth2}{RGB}{215,140,140}  
\definecolor{depth3}{RGB}{245,180,180} 
\definecolor{depth4}{RGB}{250,210,210} 
\definecolor{depth5}{RGB}{252,240,240} 

\begin{table}[htbp]
\fontsize{8}{9}\selectfont
\centering
\caption{Forgetting rate comparison (\%) across different task sequences. Results show mean $\pm$ standard deviation over 3 runs. Best results in \textbf{bold}.}
\label{tab:forgetting_comparison}
\small
\resizebox{\textwidth}{!}{%
\begin{tabular}{llccccc}
\toprule
\textbf{Task Sequence} & \textbf{Method} & \textbf{CAMUS} & \textbf{BUSI} & \textbf{ISIC} & \textbf{Kvasir} & \textbf{CheX} \\
\midrule
\multirow{2}{*}{\makecell[l]{\colorbox{depth1}{Kvasir} → \colorbox{depth2}{ISIC} → \colorbox{depth3}{CheX} \\ → \colorbox{depth4}{BUSI} → \colorbox{depth5}{CAMUS}}} & sequential & \cellcolor{depth5}\textbf{21.45}$\pm$\textbf{1.2} & \cellcolor{depth4}25.96$\pm$2.1 & \cellcolor{depth2}12.14$\pm$0.8 & \cellcolor{depth1}33.83$\pm$1.5 & \cellcolor{depth3}46.93$\pm$2.3 \\
 & PA-EWC & \cellcolor{depth5}21.80$\pm$0.9 & \cellcolor{depth4}\textbf{25.56}$\pm$\textbf{1.8} & \cellcolor{depth2}\textbf{11.23}$\pm$\textbf{0.7} & \cellcolor{depth1}\textbf{26.90}$\pm$\textbf{1.1} & \cellcolor{depth3}\textbf{43.77}$\pm$\textbf{1.9} \\
\midrule
\multirow{2}{*}{\makecell[l]{\colorbox{depth1}{CheX} → \colorbox{depth2}{CAMUS} → \colorbox{depth3}{BUSI} \\ → \colorbox{depth4}{ISIC} → \colorbox{depth5}{Kvasir}}} & sequential & \cellcolor{depth2}45.89$\pm$3.2 & \cellcolor{depth3}29.78$\pm$2.1 & \cellcolor{depth4}\textbf{8.02}$\pm$\textbf{0.5} & \cellcolor{depth5}18.34$\pm$1.4 & \cellcolor{depth1}37.92$\pm$2.6 \\
 & PA-EWC & \cellcolor{depth2}\textbf{40.26}$\pm$\textbf{2.1} & \cellcolor{depth3}\textbf{27.02}$\pm$\textbf{1.6} & \cellcolor{depth4}8.26$\pm$0.6 & \cellcolor{depth5}\textbf{15.59}$\pm$\textbf{1.0} & \cellcolor{depth1}\textbf{22.65}$\pm$\textbf{1.4} \\
\midrule
\multirow{2}{*}{\makecell[l]{\colorbox{depth1}{CAMUS} → \colorbox{depth2}{Kvasir} → \colorbox{depth3}{CheX} \\  →  \colorbox{depth4}{BUSI} → \colorbox{depth5}{ISIC}}} & sequential & \cellcolor{depth1}58.74$\pm$3.8 & \cellcolor{depth4}25.67$\pm$1.9 & \cellcolor{depth5}\textbf{7.37}$\pm$\textbf{0.4} & \cellcolor{depth2}24.12$\pm$1.6 & \cellcolor{depth3}26.95$\pm$2.3 \\
 & PA-EWC & \cellcolor{depth1}\textbf{50.30}$\pm$\textbf{2.6} & \cellcolor{depth4}\textbf{22.00}$\pm$\textbf{1.4} & \cellcolor{depth5}7.46$\pm$0.5 & \cellcolor{depth2}\textbf{20.90}$\pm$\textbf{1.2} & \cellcolor{depth3}\textbf{23.02}$\pm$\textbf{1.7} \\
\midrule
\multirow{2}{*}{\makecell[l]{\colorbox{depth1}{ISIC} → \colorbox{depth2}{CAMUS} → \colorbox{depth3}{Kvasir} \\ → \colorbox{depth4}{CheX} → \colorbox{depth5}{BUSI}}} & sequential & \cellcolor{depth2}47.56$\pm$2.9 & \cellcolor{depth5}\textbf{20.62}$\pm$\textbf{1.3} & \cellcolor{depth1}12.18$\pm$0.9 & \cellcolor{depth3}23.45$\pm$1.5 & \cellcolor{depth4}49.32$\pm$3.6 \\
 & PA-EWC & \cellcolor{depth2}\textbf{31.50}$\pm$\textbf{1.8} & \cellcolor{depth5}22.47$\pm$1.5 & \cellcolor{depth1}\textbf{9.52}$\pm$\textbf{0.6} & \cellcolor{depth3}\textbf{19.92}$\pm$\textbf{1.1} & \cellcolor{depth4}\textbf{31.74}$\pm$\textbf{2.1} \\
\midrule
\multirow{2}{*}{\makecell[l]{\colorbox{depth1}{CheX} → \colorbox{depth2}{Kvasir} → \colorbox{depth3}{CAMUS}\\ →  \colorbox{depth4}{ISIC} → \colorbox{depth5}{BUSI}}} & sequential & \cellcolor{depth3}32.69$\pm$2.0 & \cellcolor{depth5}21.97$\pm$1.4 & \cellcolor{depth4}\textbf{8.37}$\pm$\textbf{0.6} & \cellcolor{depth2}21.08$\pm$1.2 & \cellcolor{depth1}41.42$\pm$2.9 \\
 & PA-EWC & \cellcolor{depth3}\textbf{32.69}$\pm$\textbf{1.7} & \cellcolor{depth5}\textbf{19.90}$\pm$\textbf{1.1} & \cellcolor{depth4}8.57$\pm$0.7 & \cellcolor{depth2}\textbf{19.61}$\pm$\textbf{1.0} & \cellcolor{depth1}\textbf{31.78}$\pm$\textbf{2.2} \\
\bottomrule
\end{tabular}%
}
\textbf{Color Legend:} \colorbox{depth1}{1st Task} \colorbox{depth2}{2nd Task} \colorbox{depth3}{3rd Task} \colorbox{depth4}{4th Task} \colorbox{depth5}{5th Task}
\end{table}

\subsection{Typical Continual Learning Performance}

Table~\ref{tab:continual_results} presents segmentation performance across five medical imaging tasks in a continual learning setting, where tasks are learned sequentially: Kvasir → ISIC → CheXlocalize → BUSI → CAMUS. The CLIPSeg (Individual) results represent the upper bound performance when each task is trained separately with dedicated models, providing a reference for comparison. The sequential baseline clearly demonstrates the existence of catastrophic forgetting in continual learning, with substantial performance drops compared to individual training (e.g., Kvasir drops from 89.51\% to 67.36\%).
Our PA-EWC significantly outperforms existing continual learning methods. PA-EWC consistently outperforms baseline approaches, achieving the best results on three out of five tasks including Kvasir (73.42\%), CheXlocalize (56.78\%), and BUSI (71.92\%). The substantial improvements over General EWC across most datasets demonstrate PA-EWC's effective knowledge retention across diverse medical imaging modalities.

\begin{table}[htbp]
\fontsize{8}{9}\selectfont
\centering
\caption{Continual learning results on medical vision-language segmentation tasks. Performance measured by Dice coefficient (\%). Results show mean $\pm$ standard deviation over 3 runs. Best results in \textbf{bold}.}
\label{tab:continual_results}
\begin{tabular*}{\textwidth}{@{\extracolsep{\fill}}lccccc@{}}
\toprule
\textbf{Method} & \colorbox{depth1}{\textbf{Kvasir}} & \colorbox{depth2}{\textbf{ISIC}} & \colorbox{depth3}{\textbf{CheX.}} & \colorbox{depth4}{\textbf{BUSI}} & \colorbox{depth5}{\textbf{CAMUS}} \\
\midrule
\multicolumn{6}{l}{\textit{Upper Bound (Individual Task Training)}} \\
CLIPSeg (Individual) & 89.51$\pm$0.8 & 92.12$\pm$0.5 & 59.56$\pm$1.1 & 64.32$\pm$0.9 & 88.85$\pm$0.6 \\
\midrule
\multicolumn{6}{l}{\textit{Continual Learning }} \\
Sequential Baseline & 67.36$\pm$1.2 & 88.28$\pm$0.4 & 52.48$\pm$0.6 & 69.06$\pm$1.0 & 80.38$\pm$1.8 \\
General EWC & 59.31$\pm$2.1 & 83.67$\pm$1.3 & 33.85$\pm$1.8 & 68.47$\pm$2.2 & \textbf{82.11}$\pm$\textbf{1.2} \\
Self-adaptive EWC & 70.94$\pm$0.9 & \textbf{89.15}$\pm$\textbf{0.7} & 43.71$\pm$1.4 & 69.18$\pm$1.6 & 81.63$\pm$0.8 \\
PA-EWC (Ours) & \textbf{73.42}$\pm$\textbf{0.6} & 88.93$\pm$0.8 & \textbf{56.78}$\pm$\textbf{0.9} & \textbf{71.92$\pm$1.3} & 78.64$\pm$1.1 \\
\bottomrule
\end{tabular*}
\textbf{Color Legend:} \colorbox{depth1}{1st Task} \colorbox{depth2}{2nd Task} \colorbox{depth3}{3rd Task} \colorbox{depth4}{4th Task} \colorbox{depth5}{5th Task}
\end{table}

\subsection{Prompt Strategy Evaluation}

\begin{figure}[H]  
\centering
\includegraphics[width=\textwidth]{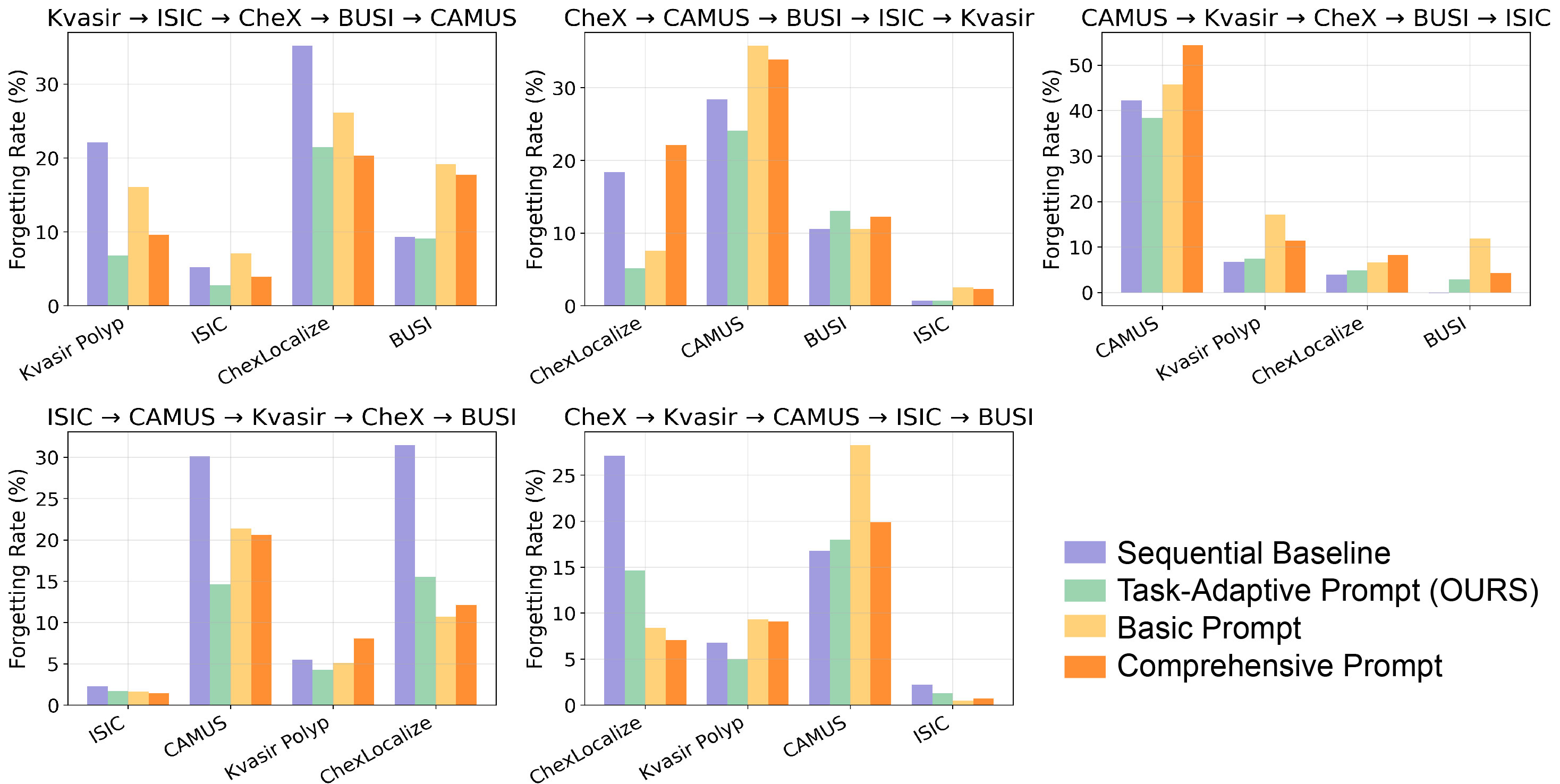}
\caption{Comparison of forgetting rates across different prompt strategies. Lower values indicate better performance with reduced catastrophic forgetting.}
\label{fig:prompt_results}
\end{figure}

Figure~\ref{fig:prompt_results} shows the forgetting rates across different prompt strategies in continual learning scenarios.. We evaluate four distinct prompt types: Sequential Baseline, Task-Adaptive Prompt (OURS), Basic Prompt, and Comprehensive Prompt. These strategies are tested across five comprehensive evaluation metrics that span different task orderings and continual learning challenges.
The Task-Adaptive Prompt strategy achieves the best overall performance with the lowest average forgetting rate (10.6\%). It consistently outperforms other methods across most evaluation metrics, demonstrating superior stability in continual learning scenarios. The Sequential Baseline approach shows the highest forgetting rates (15.2\% average), while Basic Prompt (14.6\%) and Comprehensive Prompt (14.0\%) achieve moderate performance. The results indicate that our task-adaptive approach significantly outperforms other prompts across all tasks.

\begin{figure}[H]
\centering
\includegraphics[width=\textwidth]{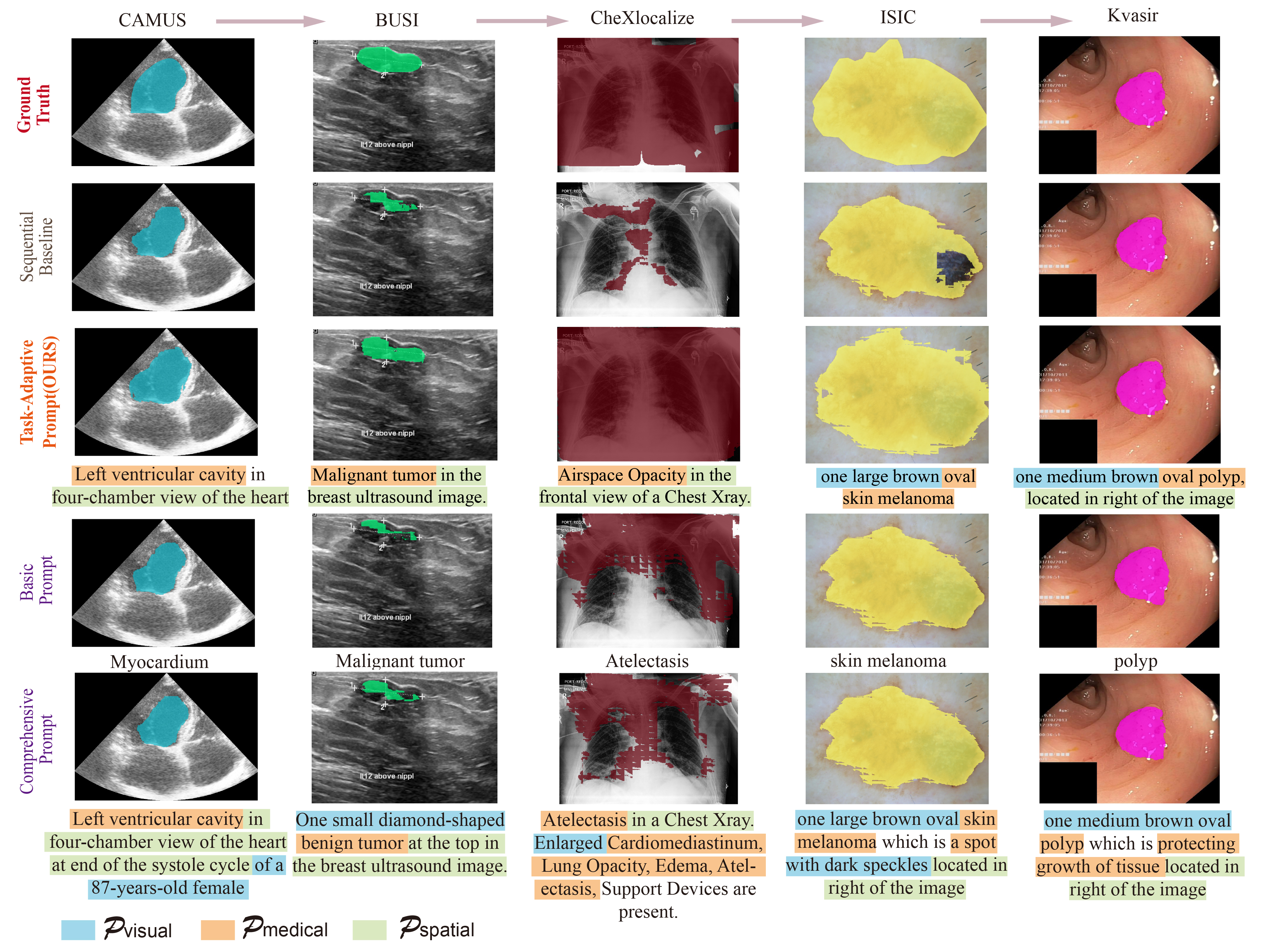}
\caption{Visualizations of segmentation results across different prompt strategies and evaluation metrics. }
\label{fig:organ}
\end{figure}

\section{Conclusion}

We introduced Prompt-Aware Adaptive Elastic Weight Consolidation (PA-EWC), a novel continual learning approach that addresses catastrophic forgetting in medical vision-language models through prompt-guided parameter specialization. Our method achieves superior performance by selectively protecting parameters based on their functional roles in processing visual, spatial, and medical semantic information. Experiments across five medical imaging datasets demonstrate that PA-EWC reduces catastrophic forgetting by up to 17.58\% while maintaining competitive computational efficiency. The approach enables medical AI systems to continuously adapt to new clinical requirements without compromising existing diagnostic capabilities, offering significant potential for real-world clinical deployment scenarios.


\clearpage
\begin{thebibliography}{99}








































\bibitem{aljundi2018memory}
Aljundi, R., Babiloni, F., Elhoseiny, M., Rohrbach, M., Tuytelaars, T.: Memory aware synapses: Learning what (not) to forget. In: Proceedings of the European Conference on Computer Vision. pp. 139--154 (2018)

\bibitem{al2020dataset}
Al-Dhabyani, W., Gomaa, M., Khaled, H., Fahmy, A.: Dataset of breast ultrasound images. Data in Brief \textbf{28}, 104863 (2020)


\bibitem{codella2019skin}
Codella, N., Rotemberg, V., Tschandl, P., Celebi, M.E., Dusza, S., Gutman, D., Helba, B., Kalloo, A., Liopyris, K., Marchetti, M., et al.: Skin lesion analysis toward melanoma detection 2018: A challenge hosted by the international skin imaging collaboration (ISIC). arXiv preprint arXiv:1902.03368 (2019)

\bibitem{esteva2017dermatologist}
Esteva, A., Kuprel, B., Novoa, R.A., Ko, J., Swetter, S.M., Blau, H.M., Thrun, S.: Dermatologist-level classification of skin cancer with deep neural networks. Nature \textbf{542}(7639), 115--118 (2017)

\bibitem{finn2017model}
Finn, C., Abbeel, P., Levine, S.: Model-agnostic meta-learning for fast adaptation of deep networks. In: International Conference on Machine Learning. pp. 1126--1135. PMLR (2017)

\bibitem{gulshan2016development}
Gulshan, V., Peng, L., Coram, M., Stumpe, M.C., Wu, D., Narayanaswamy, A., Venugopalan, S., Widner, K., Madams, T., Cuadros, J., et al.: Development and validation of a deep learning algorithm for detection of diabetic retinopathy in retinal fundus photographs. JAMA \textbf{316}(22), 2402--2410 (2016)

\bibitem{jha2020kvasir}
Jha, D., Smedsrud, P.H., Riegler, M.A., Halvorsen, P., de Lange, T., Johansen, D., Johansen, H.D.: Kvasir-SEG: A segmented polyp dataset. In: International Conference on Multimedia Modeling, pp. 451--462. Springer (2020)

\bibitem{jia2022visual}
Jia, M., Tang, L., Chen, B.C., Cardie, C., Belongie, S., Hariharan, B., Lim, S.N.: Visual prompt tuning. In: European Conference on Computer Vision. pp. 709--727. Springer (2022)

\bibitem{kirkpatrick2017overcoming}
Kirkpatrick, J., Pascanu, R., Rabinowitz, N., Veness, J., Desjardins, G., Rusu, A.A., Milan, K., Quan, J., Ramalho, T., Grabska-Barwinska, A., Hassabis, D., Clopath, C., Kumaran, D., Hadsell, R.: Overcoming catastrophic forgetting in neural networks. Proceedings of the National Academy of Sciences \textbf{114}(13), 3521--3526 (2017)

\bibitem{leclerc2019deep}
Leclerc, S., Smistad, E., Pedrosa, J., Østvik, A., Cervenansky, F., Espinosa, F., Espeland, T., Berg, E.A.R., Jodoin, P.M., Grenier, T., et al.: Deep learning for segmentation using an open large-scale dataset in 2D echocardiography. IEEE Transactions on Medical Imaging \textbf{38}(9), 2198--2210 (2019)

\bibitem{luddecke2022image}
Lüddecke, T., Ecker, A.: Image segmentation using text and image prompts. In: IEEE/CVF Conference on Computer Vision and Pattern Recognition (CVPR), pp. 7086--7096. IEEE, New Orleans (2022)

\bibitem{mallya2018packnet}
Mallya, A., Lazebnik, S.: PackNet: Adding multiple tasks to a single network by iterative pruning. In: Proceedings of the IEEE Conference on Computer Vision and Pattern Recognition. pp. 7765--7773 (2018)

\bibitem{mckinney2020international}
McKinney, S.M., Sieniek, M., Godbole, V., Godwin, J., Antropova, N., Ashrafian, H., Back, T., Chesus, M., Corrado, G.S., Darzi, A., et al.: International evaluation of an AI system for breast cancer screening. Nature \textbf{577}(7788), 89--94 (2020)

\bibitem{rajpurkar2017chexnet}
Rajpurkar, P., Irvin, J., Zhu, K., Yang, B., Mehta, H., Duan, T., Ding, D., Bagul, A., Langlotz, C., Shpanskaya, K., et al.: CheXNet: Radiologist-level pneumonia detection on chest X-rays with deep learning. arXiv preprint arXiv:1711.05225 (2017)

\bibitem{rusu2016progressive}
Rusu, A.A., Rabinowitz, N.C., Desjardins, G., Soyer, H., Kirkpatrick, J., Kavukcuoglu, K., Pascanu, R., Hadsell, R.: Progressive neural networks. arXiv preprint arXiv:1606.04671 (2016)

\bibitem{selvan2020chexlocalize}
Selvan, R., Dam, E.B., Detlefsen, N.S., Rischel, S., Sheng, K., Nielsen, M., Pai, A.: Lung segmentation from chest X-rays using variational data imputation. arXiv preprint arXiv:2005.10052 (2020)

\bibitem{smith2023coda}
Smith, J., Karlinsky, L., Gutta, V., Cascante-Bonilla, P., Kim, D., Arbelle, A., Panda, R., Feris, R., Kira, Z.: CODA-Prompt: COntinual Decomposed Attention-based Prompting for Rehearsal-Free Continual Learning. In: Proceedings of the IEEE/CVF Conference on Computer Vision and Pattern Recognition. pp. 11909--11919 (2023)

\bibitem{titsias2019functional}
Titsias, M.K., Schwarz, J., Matthews, A.G.D.G., Pascanu, R., Teh, Y.W.: Functional regularisation for continual learning with Gaussian processes. In: International Conference on Learning Representations (2019)


\bibitem{wang2022dualprompt}
Wang, Z., Zhang, Z., Ebrahimi, S., Sun, R., Zhang, H., Lee, C.Y., Ren, X., Su, G., Perot, V., Dy, J., et al.: DualPrompt: Complementary prompting for rehearsal-free continual learning. In: European Conference on Computer Vision. pp. 631--648. Springer (2022)

\bibitem{wang2022learning}
Wang, Z., Zhang, Z., Lee, C.Y., Zhang, H., Sun, R., Ren, X., Su, G., Perot, V., Dy, J., Pfister, T.: Learning to prompt for continual learning. In: Proceedings of the IEEE/CVF Conference on Computer Vision and Pattern Recognition. pp. 139--149 (2022)

\bibitem{zenke2017continual}
Zenke, F., Poole, B., Ganguli, S.: Continual learning through synaptic intelligence. In: International Conference on Machine Learning. pp. 3987--3995. PMLR (2017)

\bibitem{lopez2017gradient}
Lopez-Paz, D., Ranzato, M.: Gradient episodic memory for continual learning. In: Advances in Neural Information Processing Systems. pp. 6467--6476 (2017)

\bibitem{li2017learning}
Li, Z., Hoiem, D.: Learning without forgetting. IEEE Transactions on Pattern Analysis and Machine Intelligence \textbf{39}(12), 2935--2947 (2017)

\bibitem{buzzega2020dark}
Buzzega, P., Boschini, M., Porrello, A., Abati, D., Calderara, S.: Dark experience for general continual learning: a strong, simple baseline. In: Advances in Neural Information Processing Systems. pp. 15920--15930 (2020)

\bibitem{hu2021lora}
Hu, E.J., Shen, Y., Wallis, P., Allen-Zhu, Z., Li, Y., Wang, S., Wang, L., Chen, W.: LoRA: Low-rank adaptation of large language models. In: International Conference on Learning Representations (2022)

\bibitem{gao2021clip}
Gao, P., Geng, S., Zhang, R., Ma, T., Fang, R., Zhang, Y., Li, H., Qiao, Y.: CLIP-Adapter: Better vision-language models with feature adapters. International Journal of Computer Vision pp. 1--15 (2023)

\bibitem{zheng2023preventing}
Zheng, Z., Ma, M., Wang, K., Qin, Z., Yue, X., You, Y.: Preventing zero-shot transfer degradation in continual learning of vision-language models. In: Proceedings of the IEEE/CVF International Conference on Computer Vision. pp. 19248--19257 (2023)

\bibitem{yu2024boosting}
Yu, J., Zhuge, Y., Zhang, L., Hu, P., Wang, D., Lu, H., He, Y.: Boosting continual learning of vision-language models via mixture-of-experts adapters. In: Proceedings of the IEEE/CVF Conference on Computer Vision and Pattern Recognition. pp. 11828--11837 (2024)

\end{thebibliography}
\end{document}